\documentclass[]{pasj01}
\Received{}%{yyyy/mm/dd}
\Accepted{}%{yyyy/mm/dd}
\usepackage[separate-uncertainty=true]{siunitx}
\usepackage{lineno}
\usepackage[]{natbib}
\setcitestyle{authoryear,open={(},close={)}} %Citation-related commands

\begin{document} 

\title{ 
% \LETTERLABEL %%% <-- uncomment for LETTER article  
%\REVIEWLABEL %%% <-- uncomment for REVIEW article  
% Warm dense gas at the center of the rotating disk in a dusty starburst galaxy at z=6}
Central concentration of warm and dense molecular gas in a strongly lensed submillimeter galaxy at z=6}
%%% begin:list of authors
% Do NOT capitalize all letters in "textsc".
\author{Akiyoshi \textsc{Tsujita}\altaffilmark{1}%
% \thanks{Example: Present Address is xxxxxxxxxx}
}
\altaffiltext{1}{Institute of Astronomy, Graduate School of Science, The University of Tokyo, 2-21-1 Osawa, Mitaka, Tokyo 181-0015, Japan}
\email{tsujita@ioa.s.u-tokyo.ac.jp}

\author{Ken-ichi \textsc{Tadaki},\altaffilmark{2,3}}
\altaffiltext{2}{National Astronomical Observatory of Japan, 2-21-1 Osawa, Mitaka, Tokyo 181-8588, Japan}
\altaffiltext{3}{Department of Astronomical Science, SOKENDAI (The Graduate University for Advanced Studies), Mitaka, Tokyo 181-8588, Japan}

\author{Kotaro \textsc{Kohno}\altaffilmark{1,4}}
\altaffiltext{4}{Research Center for the Early Universe, Graduate School of Science, The University of Tokyo, 7-3-1 Hongo, Bunkyo-ku, Tokyo 113-0033, Japan}

\author{Bunyo \textsc{Hatsukade}\altaffilmark{1}}
\author{Fumi \textsc{Egusa}\altaffilmark{1}}
\author{Yoichi \textsc{Tamura}\altaffilmark{5}}
\altaffiltext{5}{Department of Physics, Graduate School of Science, Nagoya University, Furocho, Chikusa, Nagoya 464-8602, Japan}
\author{Yuri \textsc{Nishimura}\altaffilmark{1}}
\author{Jorge \textsc{Zavala}\altaffilmark{2}}
\author{Toshiki \textsc{Saito}\altaffilmark{2,6}}
\altaffiltext{6}{Department of Physics, General Studies, College of Engineering, Nihon University, 1 Nakagawara, Tokusada, Tamuramachi, Koriyama, Fukushima 963-8642, Japan}
\author{Hideki \textsc{Umehata}\altaffilmark{7,5}}
\altaffiltext{7}{Institute for Advanced Research, Nagoya University, Furocho, Chikusa, Nagoya 464-8602, Japan}
\author{Minju M. \textsc{Lee}\altaffilmark{8,9}}
\altaffiltext{8}{Cosmic Dawn Center (DAWN), Jagtvej 128, DK-2200 Copenhagen N, Denmark}
\altaffiltext{9}{DTU-Space, Technical University of Denmark, Elektrovej 327, DK2800 Kgs. Lyngby,
Denmark}
%%% end:list of authors

%% `\KeyWords{}' always has to be placed before ``\maketitle'' 
%%  List of Key Words:  https://academic.oup.com/pasj/pages/Pasj_Keywords 

\maketitle

\begin{abstract}
% Please read ``IMPORTANT NOTICE'' carefully before preparing a manuscript. 
We report the detection of the CO(12-11) line emission toward G09-83808 (or H-ATLAS J090045.4+004125),  a strongly-lensed submillimeter  galaxy at $z = 6.02$,  with Atacama Large Millimeter/submillimeter Array observations.
Combining previously detected [O\,\emissiontype{III}]\,$88\:\micron$, [N\,\emissiontype{II}]\,$205\:\micron$, and dust continuum at 0.6$\:$mm and 1.5$\:$mm, we  investigate the physical properties of the multi-phase interstellar medium in G09-83808. 
A source-plane reconstruction reveals that the region of the CO(12-11)  emission is compact ($R_\mathrm{{e, CO}}=0.49^{+0.29}_{-0.19}\,\mathrm{kpc}$) and roughly coincides with that of the dust continuum.
Non-local thermodynamic equilibrium radiative transfer modeling of CO spectral-line energy distribution  reveals that most of the CO(12-11)  emission comes from a warm (kinetic temperature of $T_{\mathrm{kin}}=320\pm170\:$ K) and dense ($\log(n_{\mathrm{H2}}/\mathrm{cm^{-3}})=5.4\pm0.6$) gas,  indicating that  the warm and dense molecular gas is concentrated in the central 0.5-kpc region. 
% We report Atacama Large Millimeter/submillimeter Array observations of [O\,\emissiontype{III}]\,$88\:\micron$, [N\,\emissiontype{II}]\,$205\:\micron$, CO(12-11) emission lines and the dust continuum at 0.6$\:$mm and 1.5$\:$mm toward G09-83808 (or H-ATLAS J090045.4+004125), a strongly lensed submillimeter bright galaxy at $z=6.03$.
% Our lens modeling indicates that CO(12-11) emitting region is as compact as that of dust continuum (effective radius of $R_\mathrm{{e, CO}}=0.49^{+0.27}_{-0.18}\,\mathrm{kpc}$). 
% Non-local thermodynamic equilibrium radiative transfer modeling with CO spectral line energy distribution indicates that most of the CO(12-11) luminosity is emitted from a warm (kinetic temperature of $T_{\mathrm{kin}}=320\pm170$ $\,$ K) and dense ($\log(n_{\mathrm{H2}}/\mathrm{cm^{-3}})=5.4\pm0.6$) gas, which means a central 0.5$\:$kpc concentration of warm gas. 
The  luminosity ratio  in G09-83808 is estimated to be $L_\mathrm{{CO(12-11)}} / L_\mathrm{{CO(6-5)}}=1.1\pm0.2$. The high ratio  is consistent with those in local active galactic nuclei (AGNs) and $6<z<7$ quasars,  the fact of which implies that G09-83808 would be a good target to explore dust-obscured AGNs in the epoch of reionization. 
In the reconstructed [O\,\emissiontype{III}]\,$88\:\micron$ and [N\,\emissiontype{II}]\,$205\:\micron$ cubes, we also find that a monotonic velocity gradient is extending over the central starburst region by a factor of two and that star-forming sub-components exist. High-resolution observations of bright [C\,\emissiontype{II}]\,$158\:\micron$ line emissions will enable us to characterize the kinematics of a possible rotating disk and the nature of the sub-components.
% ionized gas  has a monotonic velocity gradient and that  the intensity profile is well modeled  with an exponential disk with a size of $R_e\sim1.1\:$kpc. 
% These findings suggest  the existence of a rotating disk extending over the central starburst region ($R_\mathrm{{e, dust}} = 0.64\;$kpc) by a factor of two. 

% In addition, we found ionized gas clumps in the direction perpendicular to the major axis of the disk rotation.
% Our kinematic modeling shows a monotonic velocity gradient presents in the [O\,\emissiontype{III}]\,$88\:\micron$ and [N\,\emissiontype{II}]\,$205\:\micron$ maps.  Together with the fact that the velocity-integrated intensity map of these lines is well modeled with an exponential disk with a size of $R_e \sim\SI{1.1}{kpc}$, this suggests the existence of a rotating disk extending about twice as far as the central starburst region.  We also found ionized gas clumps in the direction perpendicular to the major axis of disk rotation.  
% These results suggest that the galaxy is undergoing active star formation at the center of the rotating disk.
\end{abstract}

\clearpage

% \linenumbers
\section{Introduction}

%\footnote{\textcolor{blue}{冒頭の一般的コメント\ref{general_comment_intro1}参照。}}
%\footnote{\textcolor{blue}{冒頭の一般的コメント\ref{general_logics_say_result}参照。}}
%\footnote{\textcolor{blue}{冒頭の一般的コメント\ref{general_captions}参照。}}
%\footnote{\textcolor{blue}{冒頭の一般的コメント\ref{general_grammar_relpronoun}参照。}}
% In the current most popular evolutionary scenario, submillimeter bright galaxies (SMGs) are expected to evolve into compact quiescent galaxies through major mergers or gas accretion followed by feedback processes driven by AGN or star formation.  

In the current widely accepted  scenario of the galaxy evolution, the submillimeter bright galaxy (SMG)  evolves to a massive quiescent galaxy through the central starburst with the star formation rate of hundreds of $M_\odot\mathrm{yr^{-1}}$ caused by a major merger or gravitational disk instability.
% In the current widely accepted  scenario of the galaxy evolution, the submillimeter bright galaxy (SMG)  evolves to a massive quiescent galaxy through feedback processes driven by active galactic nucleus (AGN) or star formation \citep[e.g.,][]{Toft2014}.
% Recent discoveries of massive quiescent galaxies at $z\sim 4$ suggest that first massive quiescent galaxies experienced intense starburst at $z\gtrsim5$ \citep[e.g.,][]{Glazebrook2017}. 
% One of the distinctive features of quiescent galaxies is the presence of a central compact spheroidal component (effective radius of $R_e\sim 1\:\mathrm{kpc}$ at $z\sim2$; \citealt{Dokkum2008}), which is called a bulge.  
Recently, massive quiescent galaxies have been found at $z\sim 4$, suggesting that these galaxies experienced a starburst at the earlier epoch, which is estimated to be at $z\gtrsim5$ \citep[e.g.,][]{Glazebrook2017}. 

A key to understanding how these early massive quiescent galaxies are formed is the physical property of the multi-phase interstellar medium (ISM) at the early epoch.
One way to study it is to characterize the CO spectral-line energy distribution (CO SLED),  which enables us to estimate the average physical properties of the molecular gas, such as kinetic temperature, molecular density, and gas mass.  
In particular, high-$J$ ($J_\mathrm{up}\geq 10$) CO lines trace warm and dense molecular gas with an upper energy level of $\mathrm{E_{up}}>\SI{300}{K}$ and a critical density of $n_{\mathrm{crit}}\sim 10^{5-8}\,\mathrm{cm}^{-3}$.
Since the amount of globally integrated emission from high-$J$ CO lines is difficult to explain solely with star formation, a CO SLED containing bright high-$J$ lines provides information on the contribution of other high-energy heating mechanisms, such as an AGN, shock, and cosmic rays \citep[e.g.,][]{Mashian2015}.

For local objects at $z\sim 0$, high-$J$ ($J_\mathrm{{up}}\geq 10$) CO lines  have been observed mainly with the \textit{Herschel Space Observatory} or the Stratospheric Observatory for Infrared Astronomy because the lines are  not observable from the ground.   A major problem in such space observations is the difficulty in spatially resolving the targets due to the limited beam size of the instruments.   Indeed,    even the nearest ultraluminous infrared galaxy (ULIRG)  practically appears, with \textit{Herschel} observations, as a point source (Arp~220; \citealt{Rangwala2011}) because the beam size of its Spectral and Photometric Imaging Receiver ranges between a full width at half maximum  of $\approx$ \timeform{17''}–\timeform{43''}.   In contrast,  for high-$z$ objects,  high-$J$ CO lines are redshifted to a millimeter band and hence  are observable  with ground-based interferometers like ALMA.   Ground-based observations allow us to study highly-excited molecular gas  in distant galaxies with large collecting areas, sensitive receivers, and high angular resolution.

\cite{Tadaki2022} (hereafter referred to as ``Paper I'')  concluded on the basis of the compact size of the dust-emission area and the high infrared surface brightness that G09-83808, a strongly-lensed ($\mu\sim8.4$) SMG at $z=6.02$,  is likely to be  a progenitor of a massive quiescent galaxy at $z \sim 4$.  
% This source was discovered as one of the $\SI{500}{\micron}$-riser galaxies exhibiting extremely red far-infrared (FIR) colors \citep{Ivison2016}.
% The FIR colors suggest that this source is a SMG at $z>4$. 
% The follow-up observations of CO emission lines by the Redshift Search Receiver mounted on the Large Millimeter Telescope Alfonso Serrano and \SI{890}{\micron} continuum by ALMA have revealed that this source has a spectroscopic redshift of $z=6.0269\pm 0.0006$ and is strongly gravitationally amplified (magnification factor $\mu = 9.3\pm1.0$; \citealt{Zavala2018}) by a foreground galaxy at $z=0.776$, which is identified by optical spectroscopic observations with X-shooter on the 8-m Very Large Telescope \citep{Fudamoto2017}. 
% The follow-up observations have revealed that this source 
% has a spectroscopic redshift of $z=6.0269\pm 0.0006$ and 
% is strongly gravitationally amplified (magnification factor $\mu = 9.3\pm1.0$; \citealt{Zavala2018}) by a foreground galaxy at $z=0.776$ \citep{Fudamoto2017}.  
This source has a flux density of $S_{870,\text{intr}} \sim 4~\rm{mJy}$ at 870 $\mu$m which is fainter than the two other extreme SMGs with $S_{870,\text{intr}} > 10~\rm{mJy}$ ever found at $z>6$ (HFLS3 at $z=6.3$; \citealt{Riechers2013}, SPT0311-58 at $z=6.9$; \citealt{Marrone2018}) and is likely to be a member of a more common population of starburst galaxies at $z=6$.
Gravitational lensing magnifies the solid angles of background sources, enabling us to observe distant objects with increased spatial resolution.
In this paper, we report, with help of gravitational lensing magnification, the spatially-resolved detection of the CO(12-11) line and  kinematic properties of the [O\,\emissiontype{III}]\,$88\:\micron$ and [N\,\emissiontype{II}]\,$205\:\micron$ lines in G09-83808.
Throughout the paper, we adopt a cosmology with $H_0=\SI{67.7}{km.s^{-1}.Mpc^{-1}}$, $\Omega_{\textrm{m}}=0.3106$, and $\Omega_\Lambda=0.6894$ \citep{Planck}.
% \noindent IMPORTANT NOTICE\\
% 1. ``\verb|\draft|'' creates single column and double spaces format.\\
% 2. If you comment out ``\verb|\draft|'', the output will be double column
%   and single space.\\
% 3. For cross-references, the use of ``\verb|\label|, \verb|\ref|, \verb|\citep|'' 
%   and the thebibliography environment is strongly recommended. \\
% 4. Do NOT use ``\verb|\def|, \verb|\renewcommand|''.\\
% 5. Do NOT redefine commands provided by PASJ01.cls.\\

% \newpage

\section{Observations and imaging}
% Based on the following detections of CO(5-4), CO(6-5), \ce{H2O}($2_{11}-2_{02}$) emission lines, the redshift of G09-83808 was determined to be $z=6.0269\pm 0.0006$. And also continuum emission observed with ALMA revealed that this source is strongly gravitationally lensed ($\mu = 9.3\pm1.0$, \citep{Zavala2018}) by a foreground galaxy at $z=0.776$ which is detected with X-shooter/VLT observations (\citep{Fudamoto2017}).
Paper~I  reported the initial results from ALMA Band-5 ([N\,\emissiontype{II}]\,$205\:\micron$ and 1.5-mm continuum) and Band-8 ([O\,\emissiontype{III}]\,$88\:\micron$ and 0.6-mm continuum) observations of G09-83808.
The beam sizes were \timeform{0.84''}$\times$\timeform{0.77''} for the [N\,\emissiontype{II}]\,$205\:\micron$ line and \timeform{0.76''}$\times$\timeform{0.64''} for the [O\,\emissiontype{III}]\,$88\:\micron$ line.
% In addition to these emissions, we analyze CO(12-11) data from the ALMA Band 5 observations following the imaging processes of [O\,\emissiontype{III}]\,$88\:\micron$ and [N\,\emissiontype{II}]\,$205\:\micron$ data presented in Paper I.
CO(12-11) line ($\nu_{\text{rest}}=1381.9951$\,GHz) is simultaneously observed with the [N\,\emissiontype{II}]\,$205\:\micron$ line in the ALMA Band-5 observations. The details of the observations are presented in Paper I.
% We analyze the CO(12-11) data following the image-processing procedure presented in Paper I.
% At $z=6$, CO(12-11) ($\nu_\mathrm{obs}$=196.7 GHz) and [N\,\emissiontype{II}]\,$205\:\micron$ ($\nu_\mathrm{obs}$=207.9 GHz) line emissions can be simultaneously observed with ALMA Band-5 receivers.

By applying a clean mask constructed for 1.5\,mm continuum (white solid line in the data panel of figure~\ref{CO1211_modeling}; see Paper I for details), we make channel maps of the CO(12-11) line at 100$\:$km$\:$s$^{-1}$ spectral resolution with a Briggs weighting of robustness parameter of 2.0, resulting in a beam size of 0\farcs87$\times$0\farcs77 (figure \ref{CO1211_channelmaps} in appendix 1). 
% \textcolor{red}{The line spectra for [O\,\emissiontype{III}]\,$88\:\micron$, [N\,\emissiontype{II}]\,$205\:\micron$, and CO(12-11), measured within the masked region in the cleaned images are shown in figure \ref{observed_spectrum}.}

The CO(12-11) line is integrated over a velocity range of $-$250$\:$km$\:$s$^{-1}$ to +250$\:$km$\:$s$^{-1}$.  
The velocity offset is calculated relative to the systemic redshift of $z=6.0244\pm0.0003$ (see section \ref{channelmap}).
We resolve the lensed arcs of CO(12-11) emission in the integrated intensity map, where the peak line fluxes and  noise levels are $0.19\pm0.02\:\mathrm{Jy\:beam^{-1}\:km\:s^{-1}}\ (7.8\sigma)$  and $0.15\pm0.02\:\mathrm{Jy\:beam^{-1}\:km\:s^{-1}}\ (6.0\sigma)$ in the north-west and south-east arcs, respectively (the cleaned image panel of figure \ref{CO1211_modeling}).  We also show the the channel maps of [O\,\emissiontype{III}]\,$88\:\micron$ and [N\,\emissiontype{II}]\,$205\:\micron$ emissions with peak signal-to-noise ratio (SNR) of  larger than 4 in the cleaned image panels of figure~\ref{OIII_channel_modeling_1comp}, \ref{NII_channel_modeling_1comp}. The noise level is $1\sigma=0.45\;$mJy/beam per $100\;\mathrm{km\;s^{-1}}$ in the [O\,\emissiontype{III}]\,$88\:\micron$ cube,  $1\sigma=0.13\;$mJy/beam per $100\;\mathrm{km\;s^{-1}}$ in the [N\,\emissiontype{II}]\,$205\:\micron$ cube.

\section{Lens modeling}
We use the open-source code \texttt{GLAFIC} \citep{Oguri2010} to quantify the gravitational lensing effect and to reconstruct the source spatial structure  with a parametric approach in the cleaned image plane. 
We assume that the mass distribution of the foreground lens galaxy, which is situated at $z=0.776$ \citep{Fudamoto2017}, follows a singular isothermal ellipsoid with external shear and that the brightness distribution of the background source galaxy at $z=6.02$ has a S$\acute{\mathrm{e}}$rsic profile.  
We first determine the lens parameters by modeling the 1.5-mm continuum map, which has the highest SNR among our ALMA data. 
We perform all lens modeling in the clean mask region (white solid line in the cleaned image panel of figure~\ref{CO1211_modeling},  \ref{OIII_channel_modeling_1comp}, \ref{NII_channel_modeling_1comp}) to focus on the emitting region and reduce the amount of calculation.
 Then, we reconstruct the background source, where all parameters of the foreground source are fixed.
The details  of the method for determining the lens parameters and for the source reconstructions in the continuum maps and  integrated intensity maps of the [O\,\emissiontype{III}]\,$88\:\micron$ and the [N\,\emissiontype{II}]\,$205\:\micron$ lines are described in Paper~I.
In this paper, we focus on source reconstruction in the integrated intensity map of the CO(12-11) line and the channel maps of the [O\,\emissiontype{III}]\,$88\:\micron$ and [N\,\emissiontype{II}]\,$205\:\micron$ lines.
% The parameters of our lens model are summarized in Table \ref{lensParameters}.

% \begin{table*}
% \tbl{Lens parameters. $x_l$ and $y_l$ are the position of the lens in arcseconds relative to the phase center, θE,L is the Einstein radius of the lens, eL is the ellipticity and φL is the position angle in degrees east of north.}{%
% \begin{tabular}{ccccccc}
% \hline\hline\noalign{\vskip3pt} 
%  & $z_l$ & $\sigma$ & $(\Delta x_l, \Delta y_l)$ & $e$ & $\theta_e$ & $r_{\mathrm{core}}$ \\ [2pt] 
% \hline\noalign{\vskip3pt} 
% SIE & 0.776 & \SI{258}{km.s^{-1}} & $(0.340", 0.175")$ & 0.101 & $\ang{78.8}$ & 0 \\
% \hline\hline\noalign{\vskip3pt} 
%  & $z_l$ & $\sigma$ & $(\Delta x_l, \Delta y_l)$ & $\gamma$ & $\theta_\gamma$ & $\kappa$ \\ [2pt] 
% \hline\noalign{\vskip3pt} 
% PERT & 0.776 & - & (0.340", 0.175") & 0.0392 & $\ang{46.7}$ & 0 \\
% \hline\noalign{\vskip3pt} 
% \end{tabular}
% }\label{lensParameters}
% \end{table*}

\subsection{Integrated intensity map of the CO(12-11) emission}
\label{co1211mom0}
Since the SNR and angular resolution of the integrated intensity map of the CO(12-11) emission are worse than those of the 1.5-mm continuum map (75$\sigma$ detection and \timeform{0.48''}$\times$\timeform{0.41''}), it is not straightforward to determine all the parameters of the background source (spatial position ($x$, $y$), total flux $Sdv$, effective radius $R_e$, major-to-minor axis ratio $q$, position angle $\theta_q$, and S$\acute{\mathrm{e}}$rsic index $n$). Thus, we made some assumptions in our modeling.
Given that the surface brightness profile of 1.5-mm continuum emission is well approximated by S$\acute{\mathrm{e}}$rsic profile of  $n=$1.06--1.30 (Paper~I), we adopt an exponential disk profile with $n=1$.
We fix the axis ratio and  position angle to $q=0.93$ and $\theta_q=108$ deg, respectively, the values of which are derived from the 1.5-mm continuum data.  Then, we fit  the integrated intensity map of the CO(12-11) emission with the remaining free parameters of $xy$-position, $Sdv$, and $R_e$.
% The best-fit model indicates that the source position of the CO(12-11) emission is almost identical to that of 1.5$\:$mm continuum emission.
% We fix the source position to that of the 1.5$\:$mm continuum and perform the fitting again in order to compare the CO(12-11) source size with that of the continuum more fairly. 

\begin{figure*}
 \begin{center}
  \includegraphics[width=1.0\linewidth]{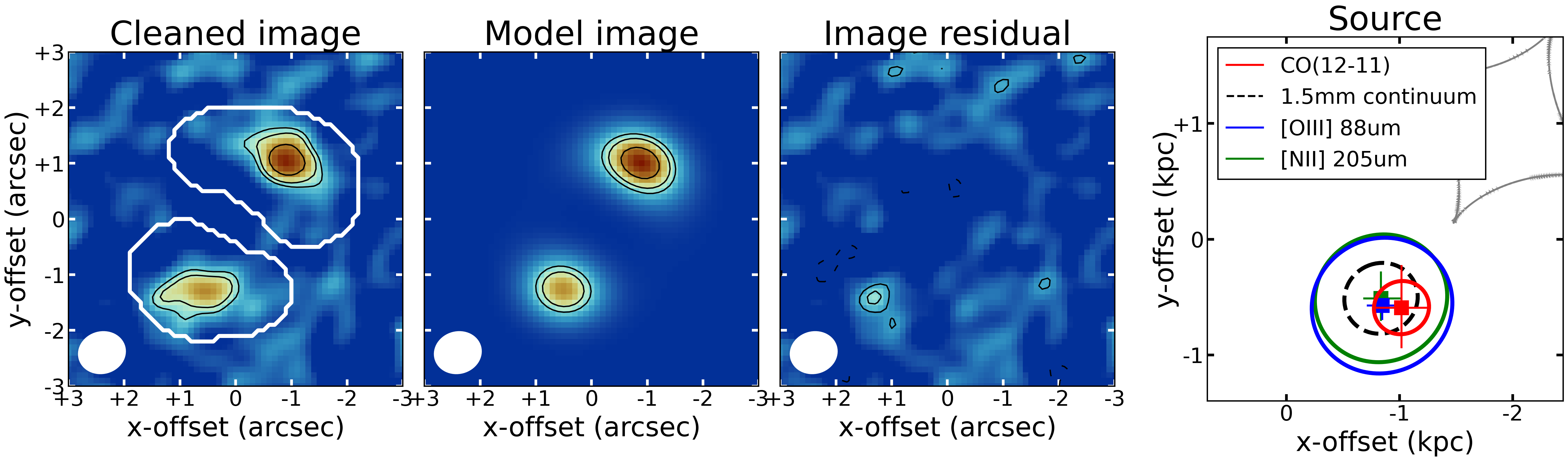}
  %\label{CO1211_modeling} % WB: moved to after the caption
 \end{center}
\caption{Source reconstruction for the CO(12-11) integrated intensity map. Panels, from left to right, show the observed data (cleaned image), obtained best-fit model image, residuals, and reconstructed source model of CO(12-11). The black contours in the cleaned image and model image panel are drawn at [3, 4, 6, 8]$\times \sigma$, where $\sigma=0.02\:\mathrm{Jy\:beam^{-1}\:km\:s^{-1}}$. Black contours in the residual panel are drawn at [-3, -2, 2, 3]$\times \sigma$.  In the  right-most panel, the reconstructed sources of the other emissions are also shown for comparison.  The white solid line in the data panel  indicates the mask region (the same as the clean mask) where the lens modeling is performed.  The synthesized beam size is  displayed in the lower-left corner of each panel. The filled squares and  encircling ellipses in the  rightmost panel indicate the positions and sizes of the sources.  The gray solid line represents the caustics of our lens model.}
\label{CO1211_modeling} % WB: moved from before the caption
\end{figure*}

Figure~\ref{CO1211_modeling} shows the cleaned image, obtained best-fit model image, residual, and reconstructed source model of CO(12-11) integrated intensity map. In the source panel, we also show the source models of the 1.5-mm continuum and integrated intensity map of [O\,\emissiontype{III}]\,$88\:\micron$ line and [N\,\emissiontype{II}]\,$205\:\micron$ line reconstructed in the same manner.
To estimate the uncertainties of the derived source parameters, we add a $1\sigma$ noise map convolved with a dirty beam to the clean image and fit the noise-added images.
Then, we take the 16th and the 84th percentile of the best-fit values from 500 Monte Carlo runs as the uncertainties.
% $R_e=\timeform{0.10''}^{+\timeform{0.02''}}_{-\timeform{0.04"}}$

The intrinsic effective radius of the CO(12-11) emission is determined to be $R_e=0.08^{+0.05}_{-0.03}\ \mathrm{arcsec}$, corresponding to $0.49^{+0.29}_{-0.19}\ \mathrm{kpc}$ at $z=6.02$, after correction of the lensing effect. 
This value is consistent with the effective radius of the dust continuum emission (\timeform{0.11''}--\timeform{0.12''}; Paper~I), but is by a factor of two smaller than those of [O\,\emissiontype{III}]\,$88\:\micron$ and [N\,\emissiontype{II}]\,$205\:\micron$ emissions (\timeform{0.20''}--\timeform{0.21''}; Paper~I). The total magnification factor, given by the ratio between the total flux densities in the image and source plane, is calculated to be $\mu_{\mathrm{CO(12-11)}}=9.3^{+7.2}_{-1.7}$, which is consistent with those of the dust continuum at 0.6$\:$mm and 1.5$\:$mm, [O\,\emissiontype{III}]\,$88\:\micron$, and [N\,\emissiontype{II}]\,$205\:\micron$ ($8.0-8.4$; Paper~I).
The position of CO(12-11) emission is consistent with that of the dust continuum, [N\,\emissiontype{II}]\,$205\:\micron$, and [O\,\emissiontype{III}]\,$88\:\micron$ within the $1\sigma$ uncertainty. In appendix 2, we provide a more detailed discussion of CO(12-11) size measurement.

According to past  high-resolution observations of other high-z SMGs, the CO emission region  is smaller  for a higher excitation level  \citep[e.g.,][]{Weiss2005, Apostolovski2019, Jarugula2021}. In addition, it is  known in high-z SMGs that the low-$J$ CO emission region is more extended than that of dust and that the distribution of the mid/high-$J$ emission is  similar to, or slightly more extended than, that of dust even at $J_{\mathrm{up}}=8,9$ \citep[e.g.,][]{Spilker2015,Chen2017, Rivela2018, Apostolovski2019, Dong2019, Rybak2020, Jarugula2021}.   
% , and is found to distribute comparable to the dust continuum. 
To our best knowledge, the CO(12-11) emission in G09-83808 is the most excited among the spatially-resolved CO rotational lines.
% These source reconstructions of the dust continuum and the integrated intensity map of CO(12-11), [O\,\emissiontype{III}]\,$88\:\micron$, and [N\,\emissiontype{II}]\,$205\:\micron$ line show that each line map is well modeled by a disk profile with a S$\acute{\mathrm{e}}$rsic index of n=1, and that the ionized gas traced by the [O\,\emissiontype{III}]\,$88\:\micron$ and [N\,\emissiontype{II}]\,$205\:\micron$ lines is more extended than the CO(12-11) line and dust continuum.  

% In the next subsection, we reconstruct velocity channel of [O\,\emissiontype{III}]\,$88\:\micron$ and [N\,\emissiontype{II}]\,$205\:\micron$ line to investigate the intrinsic kinematic information of the ionized gas.
% through the source reconstruction of each velocity channel of [O\,\emissiontype{III}]\,$88\:\micron$ and [N\,\emissiontype{II}]\,$205\:\micron$ line.

\subsection{Channel maps of the [O\,\emissiontype{III}]\,$\SI{88}~{\micron}$ and [N\,\emissiontype{II}]\,$\SI{205}~{\micron}$ emission} \label{channelmap}
\begin{figure*}
 \begin{center}
  \includegraphics[width=0.82\linewidth]{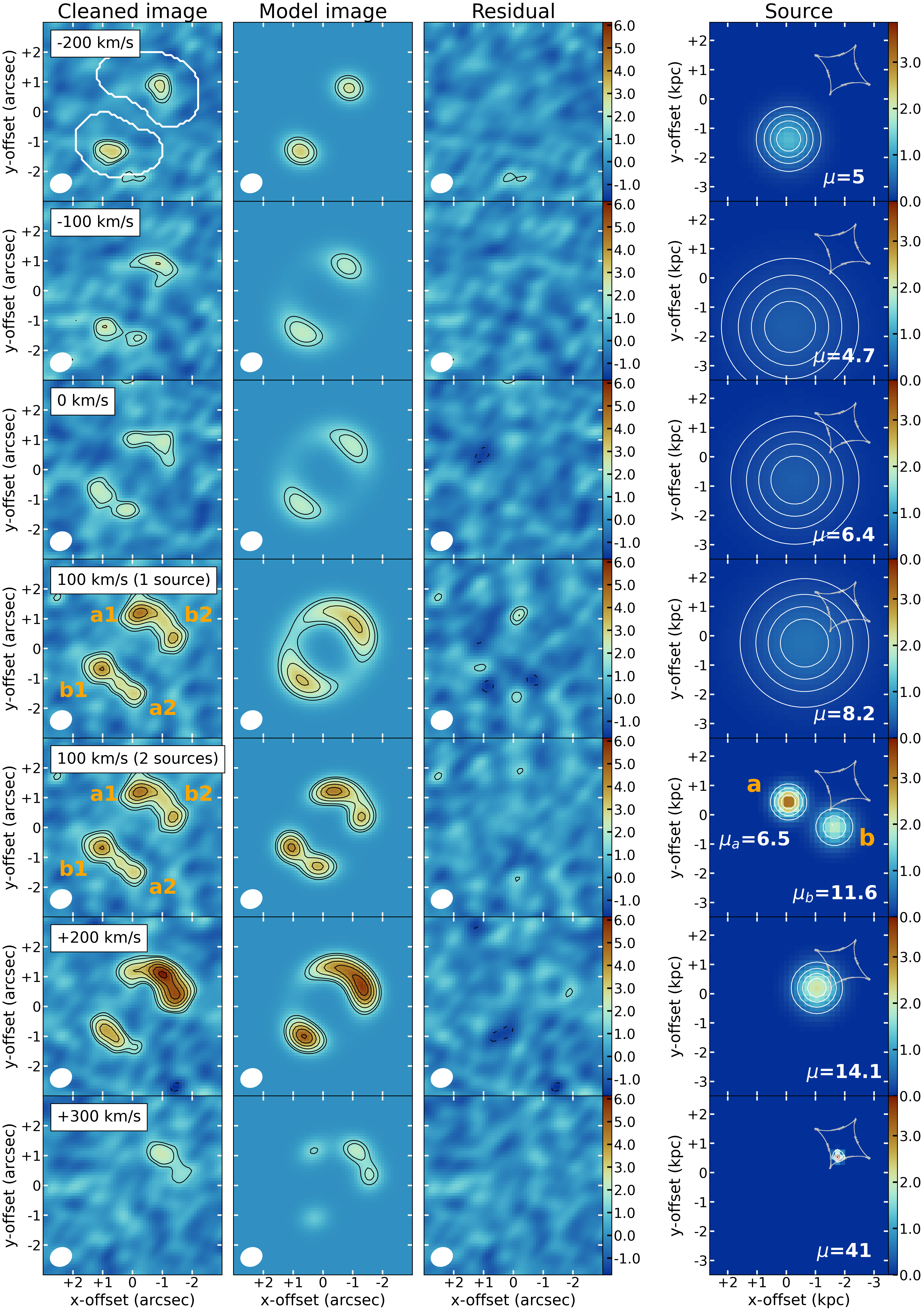}
 \end{center}
\caption{Source reconstructions for the [O\,\emissiontype{III}]\,$88\:\micron$ channel maps.  The peak SNRs of the cleaned [O\,\emissiontype{III}]\,$88\:\micron$ images are [$7.8\sigma$, $6.2\sigma$, $5.6\sigma$, $11.5\sigma$, $14.6\sigma$, $6.0\sigma$].  White solid lines on the top-left panel show the mask region (clean mask) in which the lens modeling is performed.  The black contours in the cleaned image panel and model image panel are drawn at [3, 4, 6, 8, 10, 12, 14]$\times \sigma$. The black dashed and solid contours in the residual panel are drawn at [-3, 3, 4, 5]$\times \sigma$. The white contours in the source panel show 20\%, 40\%, 60\%, and 80\% of the peak. The modeling at $\SI{0}{km.s^{-1}}$ does not fit well with a single S$\acute{\mathrm{e}}$rsic source, leaving a residual of more than $5\sigma$ (see $\SI{100}{km.s^{-1}}$ (1 source) row), while the modeling with two S$\acute{\mathrm{e}}$rsic sources fit well with $\sim 3\sigma$ residual ($\SI{100}{km.s^{-1}}$ (2 sources) row). The (a1, a2) and (b1, b2) component in the image plane corresponds to a and b source in the source plane. The gray solid diamond-shaped lines in the source panel represent the caustics of our lens model.  Colorbar units are $\si{mJy.beam^{-1}}$. The magnification factor $\mu$ of each channel is shown in the source panel. The synthesized beam size is displayed in the lower-left corner of each panel. }
 \label{OIII_channel_modeling_1comp}
\end{figure*}

\begin{figure*}
 \begin{center}
  \includegraphics[width=0.82\linewidth]{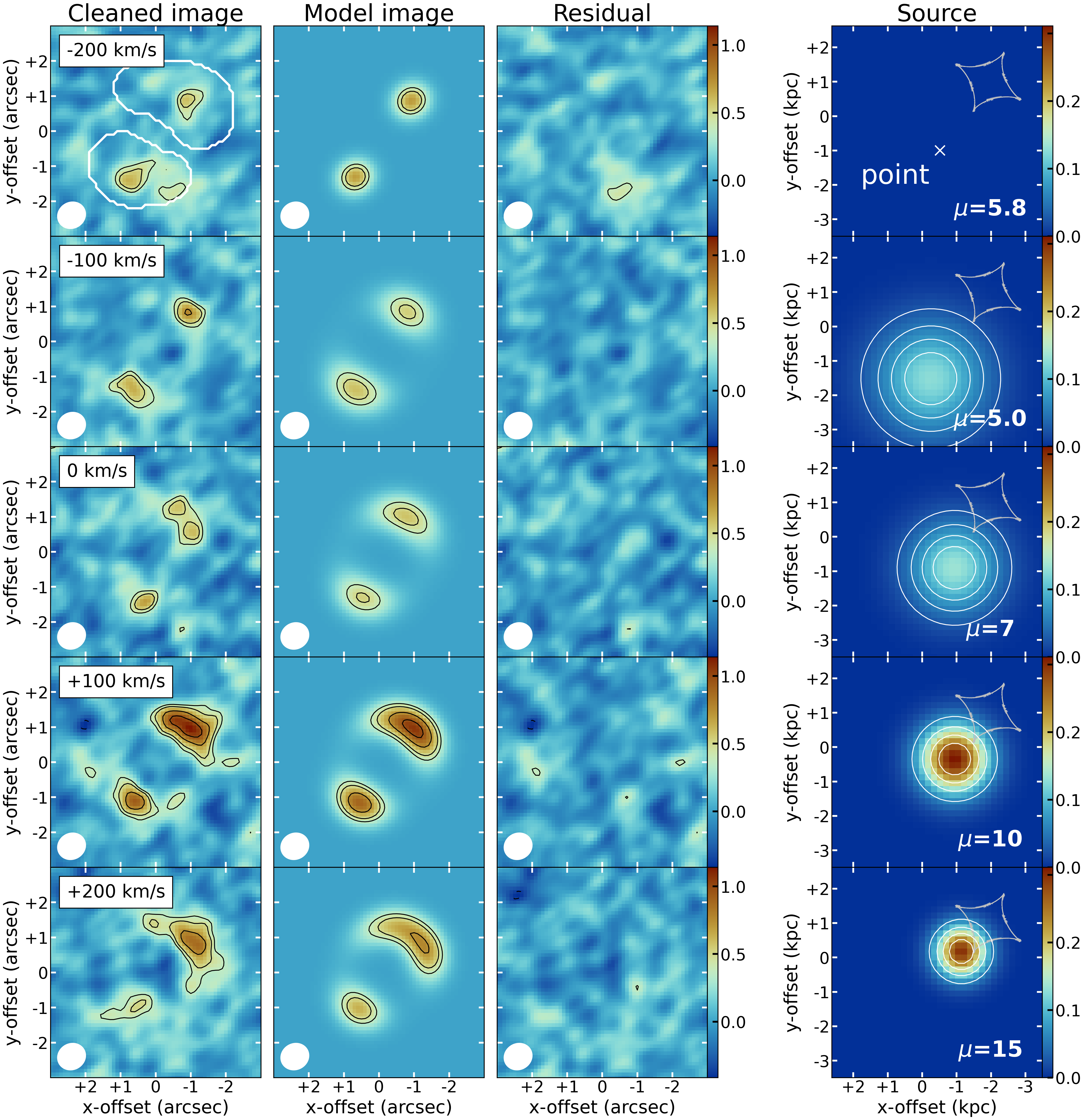}
 \end{center}
\caption{Source reconstructions for the [N\,\emissiontype{II}]\,$205\:\micron$ channel maps.  The peak SNRs of the observed data are [$4.9\sigma$, $5.9\sigma$, $5.3\sigma$, $8.8\sigma$, $7\sigma$, $4\sigma$].  The contour levels and annotations are the same as figure \ref{OIII_channel_modeling_1comp}. The modeling at $\SI{-200}{km.s^{-1}}$ is performed with a point source whose position is shown as a white cross mark in the source panel.}
 \label{NII_channel_modeling_1comp}
\end{figure*}

\begin{figure*}
 \begin{center}
  \includegraphics[width=1.0\linewidth]{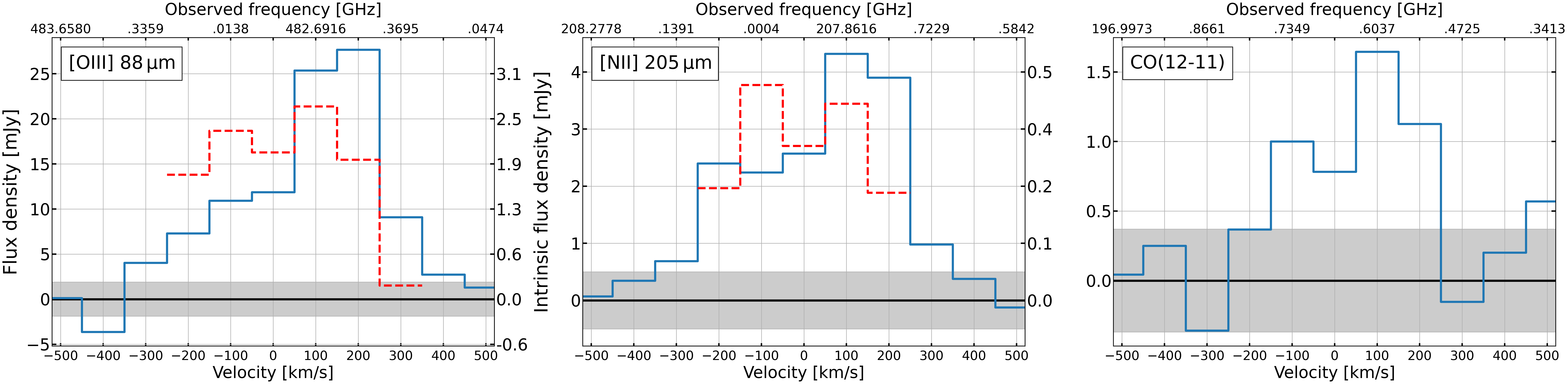}
  %\label{CO1211_modeling} % WB: moved to after the caption
 \end{center}
\caption{Integrated line spectra for [O\,\emissiontype{III}]\,$88\:\micron$, [N\,\emissiontype{II}]\,$205\:\micron$, and CO(12-11), measured within the masked region in the cleaned images (blue solid lines) and their reconstructed sources (red dashed lines). The intrinsic flux density is plotted on the right axis. To make it easier to compare the profiles, source plane line profiles are multiplied by the total magnifications. Gray shaded areas represent the $\pm 1\sigma$ uncertainties.}
\label{observed_spectrum} % WB: moved from before the caption
\end{figure*}

% For the [O\,\emissiontype{III}]\,$88\:\micron$ $\SI{100}{km.s^{-1}}$ channel maps, the [N\,\emissiontype{II}]\,$205\:\micron$ $\SI{100}{km.s^{-1}}$ channel maps, and the CO(12-11) integrated intensity map, which are discussed in this paper, the peak S/N ratios are all about $10\sigma$.  
Both doubly-ionized oxygen and ionized nitrogen have higher ionization potentials than that of neutral hydrogen and work as tracers of ionized gas.
The observed SNRs of the [O\,\emissiontype{III}]\,$\SI{88}~{\micron}$ and [N\,\emissiontype{II}]\,$\SI{205}~{\micron}$ channel maps with a channel width of 100$\:$km$\:$s$^{-1}$ are  marginally higher than those of the CO(12-11) channel maps, and we  extract  kinematic information of the ionized gas  through source reconstruction of the [O\,\emissiontype{III}]\,$\SI{88}~{\micron}$ and [N\,\emissiontype{II}]\,$\SI{205}~{\micron}$ channel maps, as follows.
% we determine the source positions of each velocity component in the [O\,\emissiontype{III}]\,$\SI{88}~{\micron}$ and [N\,\emissiontype{II}]\,$\SI{205}~{\micron}$ channel maps with a channel width of 100$\:$km$\:$s$^{-1}$.
We adopt a circular Gaussian source with S$\acute{\mathrm{e}}$rsic index $n=0.5$ rather than $n=1$ as used in the modeling of the integrated intensity maps so that the fitting is more sensitive to the peak position.
%we could fit more sensitively to the part with the peak flux, rather than the extended arc component which have lower SNR.  
% In modeling on each channel, we also perform  fitting by using a point source instead of an extended source and adopt a better model based on the reduced chi-square values. 
Since the angular resolution and SNRs are relatively low, a point source may fit better than an extended circular Gaussian source. Thus, we also perform the source reconstruction using a point source instead of a Gaussian source and adopt a better model based on their chi-square values.
As a result, the point source model is adopted only for the $\SI{-200}{km.s^{-1}}$ channel of [N\,\emissiontype{II}]\,$\SI{205}{\micron}$.
The uncertainties of the parameters are estimated in the same manner as described in section \ref{co1211mom0}.
Figure~\ref{OIII_channel_modeling_1comp}, \ref{NII_channel_modeling_1comp} show the cleaned images, obtained best-fit model images, residuals, and source images of each channel map. 
%The contours levels of the residual panels denote $[-4\sigma, -3\sigma, +3\sigma, +4\sigma]$. 
The residual maps demonstrate that the fitting with a circular Gaussian source has generally worked well, remaining $\lesssim 3\sigma$ peak residual.
However, as shown in $\SI{100}{km.s^{-1}}$ (1 source) row panels in figure \ref{OIII_channel_modeling_1comp}, a non-negligible structure with SNR $>5$ remains at the $\SI{100}{km.s^{-1}}$ channel of  [O\,\emissiontype{III}]\,$88\:\micron$ residual map using a single Gaussian source.
% However, only in the [O\,\emissiontype{III}]\,$88\:\micron$ channel map at $\SI{0}{km.s^{-1}}$, a residual emissions remain at a $>4\sigma$ level (referred to as \textbf{a1} component). 
Then, fitting using two circular Gaussian sources is performed ($\SI{100}{km.s^{-1}}$ (2 sources row) panels). The \textbf{a} and \textbf{b} sources correspond to (\textbf{a1}, \textbf{a2}) and (\textbf{b1}, \textbf{b2}) components in the image plane and reduce the residuals to $<3\sigma$.
% Fitting with two circular Gaussian sources (\textbf{a} and \textbf{b} components in figure~\ref{channel_position_1comp}) reduce the residuals to $<3\sigma$ ($\SI{0}{km.s^{-1}}$ \textcircled{\scriptsize 2} row panels).
The best fit magnification factors of each [O\,\emissiontype{III}]\,$88\:\micron$ source at $\SI{-200}{km.s^{-1}}$, $\SI{-100}{km.s^{-1}}$, $\SI{0}{km.s^{-1}}$, $\SI{100}{km.s^{-1}}$ (source a), $\SI{100}{km.s^{-1}}$ (source b), $\SI{200}{km.s^{-1}}$, and $\SI{300}{km.s^{-1}}$ are $5_{-1}^{+1}$, $4.7_{-0.8}^{+1.3}$, $6.4_{-0.6}^{+2.2}$, $6.5_{-0.4}^{+1.8}$, $11.6_{-3.3}^{+1.6}$, $14.1_{-1.4}^{+2.4}$, and $41_{-7}^{+9}$, respectively. 
The best fit magnification factors of each [N\,\emissiontype{II}]\,$205\:\micron$ source from $\SI{-200}{km.s^{-1}}$ to $\SI{+200}{km.s^{-1}}$ are $5.8_{-1.2}^{+2.5}$, $5.0_{-1.3}^{+2.7}$, $7_{-3}^{+6}$, $10_{-2}^{+6}$, $15_{-7}^{+8}$, respectively. 
Line profiles of the observed clean images measured within the masked region and the reconstructed [O\,\emissiontype{III}]\,$88\:\micron$ and [N\,\emissiontype{II}]\,$205\:\micron$ sources are shown in figure \ref{observed_spectrum}.
The redshifted parts of both emissions are closer to the caustics and more magnified than the blueshifted parts. The differential magnification causes the lensed image profiles to contain more flux at positive velocities, even though the intrinsic line profiles are more symmetrical.  
Based on the single gaussian fit to the reconstructed profile of the [O\,\emissiontype{III}]\,$88\:\micron$ line, the redshift is determined to be $z=6.0244\pm0.0003$, about $\SI{-100}{km.s^{-1}}$ shift from the previous value ($z=6.0269\pm0.0006$; \citealt{Zavala2018}).

\section{Physical properties of highly excited gas}
\begin{figure}
 \begin{center}
  \includegraphics[width=1.0\linewidth]{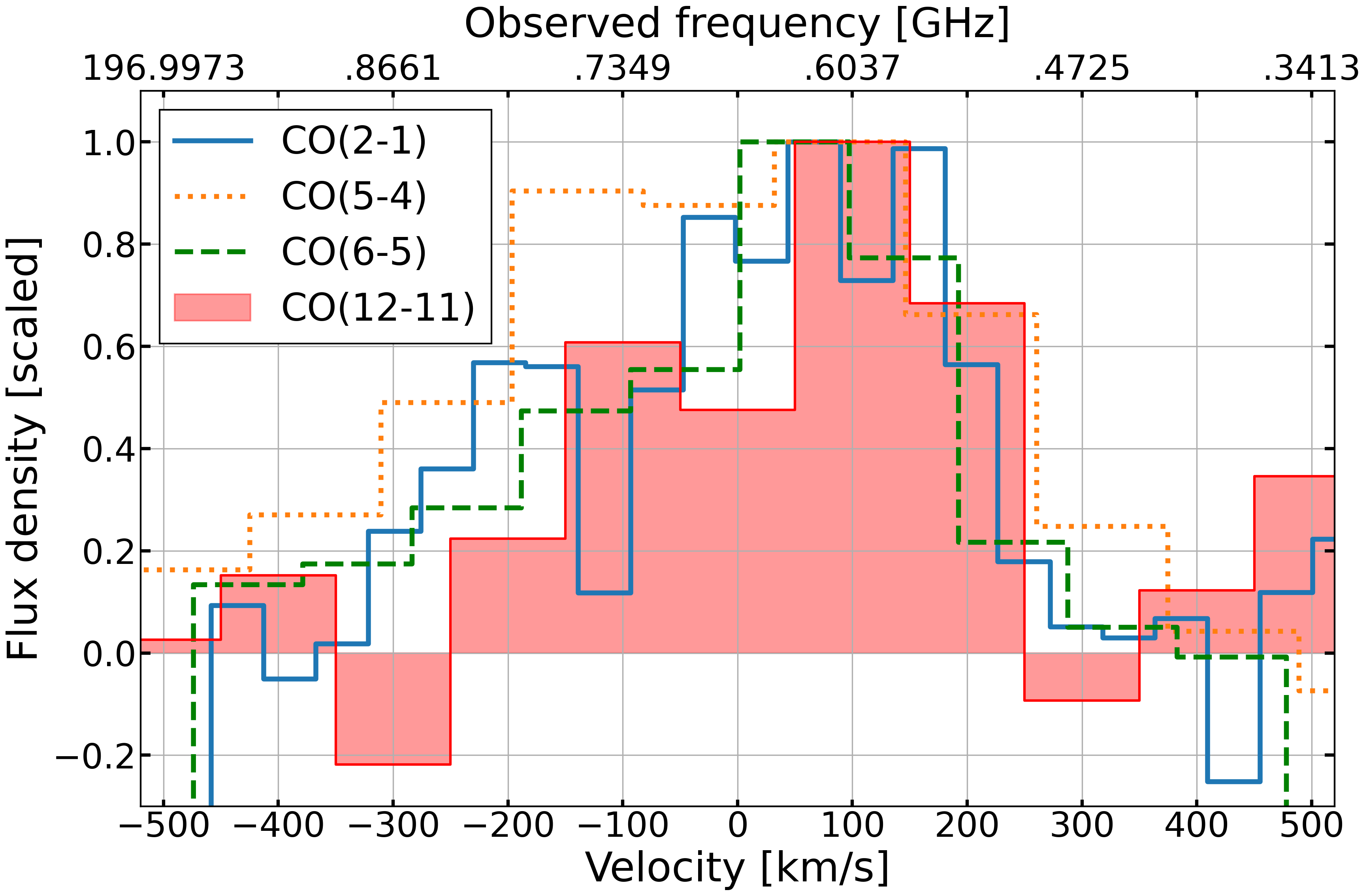}
 \end{center}
\caption{The CO(12-11) line profile measured in the cleaned image compared with those of CO(2-1) from \citet{Zavala2022}, CO(5-4), and CO(6-5) from \citet{Zavala2018}. The lines are scaled to match the line peak to compare the line profiles.}
\label{CO_spectra}
\end{figure}

The detection of CO(12-11) line (whose  upper energy level is $E_{\mathrm{up}}=\SI{431}{K}$) suggests the existence of  highly excited molecular gas.  In order to  investigate quantitatively the physical properties of the molecular gas, we model the CO SLED, using the 1D non-local thermal equilibrium (non-LTE) radiative transfer code \texttt{RADEX} \citep{vanderTak2007}.  In the modeling, we incorporate, in addition to the CO(12-11) line detected in this work, the CO(2-1) (Zavala et~al.~in prep), CO(5-4), and CO(6-5)  \citep{Fudamoto2017, Zavala2018} lines  detected in G09-83808 (Table~\ref{COdata}). 
We use the galaxy-integrated flux densities for any of the lines for fair comparison even though the CO(2-1) and CO(12-11) line distributions were spatially resolved.  In other words, the present analysis does not take into account the  dependency of the gravitational lensing effect  on the source position.
Although the differential magnification effect  could in principle affect the physical properties derived from a spatially unresolved CO SLED \citep[e.g.,][]{Dong2019, Yang2019},   we consider that its effect is  negligible in the present analysis for the following reasons. First, with our lens model, the total magnification factors of the spatially extended [O\,\emissiontype{III}]\,$88\:\micron$ and [N\,\emissiontype{II}]\,$205\:\micron$ lines are consistent with those of compact dust continuum and CO(12-11) line (section~\ref{co1211mom0}).
Therefore, the difference in the total magnification effect between the extended low-$J$ CO emission and compact high-$J$ CO emission is expected to be small.
Secondly, as shown in figure \ref{CO_spectra}, the line profiles of CO(2-1), CO(5-4), CO(6-5), and CO(12-11) do not differ significantly.
And the total magnification factors of [O\,\emissiontype{III}]\,$88\:\micron$ and [N\,\emissiontype{II}]\,$205\:\micron$ emissions which are obtained by the source reconstructions for each channel map are $8.4\pm1.1$ and $8\pm3$, respectively, which are consistent with the average magnification factors of $8.09_{-0.08}^{+0.08}$ and $8.3_{-0.2}^{+0.1}$ derived from the source reconstructions for the integrated intensity maps.
Therefore, the difference in the magnification effect due to the differences in velocity structure is also expected not to be significant.

The inputs  to \texttt{RADEX} are  a kinetic temperature ($T_{\textrm{kin}}$), molecular hydrogen density ($n_{\mathrm{H2}}$), and CO column density ($N_{\textrm{CO}}$). We assume an expanding sphere geometry and set $dV=\SI{400}{km.s^{-1}}$, which is roughly the FWHM of the CO lines, the background temperature to 19.2$\:$K, and the cosmic microwave background temperature to $z = 6.02$. We search for the minimum chi-square within the parameter space of $T_{\textrm{kin}}=20-700$\,K, $\log(n_{\mathrm{H2}}/\mathrm{cm^{-3}})=2-6.7$, $\log(N_{\mathrm{CO}}/\mathrm{cm^{-2}})=16-20.5$.   Our fitting attempt with the CO SLED with a single component results in  a solution that does  not converge within the search range. Hence, the model requires an additional component.  
% Indeed, some previous studies \citep[e.g.,][]{Wang2019} reported that  the fitting results of CO SLEDs were  markedly improved when the applied model was switched from  a single-component  one  to two-component one composed of a cold gas component  originating in star formation and a warm gas component, which is presumably directly heated by a central AGN.  
In our case of G09-83808, the number of the observed CO line data is only four; thus,   some additional constraints on the free parameters are necessary to perform two-component model-fitting and obtain meaningful results.    
A constraint that we set is that the molecular gas mass must not exceed the dynamical mass ($M_{\mathrm{dyn}}=5\times 10^{10}\,M_\odot$ including $1\sigma$ error; \citealt{Zavala2018}).   Another constraint is that the CO SLED of both the warm and cool gas components satisfy $L_\mathrm{{CO(13-12)}} < L_\mathrm{{CO(12-11)}}$.  We also assume that the temperature of the cold component is $T_{\mathrm{kin, cold}}<100\, \mathrm{K}$.   In fact,  most of the local ULIRGs, high-z quasars, and SMGs satisfy the second and third constraints \citep[e.g.,][]{Mashian2015,Li2020,Yang2017}.  We note that  the determined $T_{\mathrm{kin}}$ or $n_{\mathrm{H2}}$ would  go beyond the search ranges without these constraints in the model fitting.
% As for CO column density, one prior is that the total amount of gas producing the observed CO luminosities should be no more than the total dynamical mass of this system. This leads to
% \begin{equation}
%     N_\mathrm{CO}<\frac{M_{\mathrm{dyn}} X_{\mathrm{CO}}}{\mu m_{\mathrm{H} 2} A}\left[\mathrm{~cm}^{-2}\right]
% \end{equation}
% \noindent
% where the dynamical mass is $M_{\mathrm{dyn}}=2.6\times 10^{10}\,M_\odot$ \citep{Zavala2018}, the ratio of CO relative to \ce{H2} is $X_{\mathrm{CO}}=10^{-4}$, the mean molecular weight is $\mu=1.4$, $m_{\mathrm{H} 2}$ is the \ce{H2} molecule mass, and the source area is $A=\pi(\SI{1}{kpc})^2$. 
% We also imposed the constraints that the cold component should account for at least 60\% of $\mathrm{L_{CO(2-1)}}$ and the warm component should account for at least 60\% of $\mathrm{L_{CO(12-11)}}$, the temperature of the cold component $T_{\mathrm{kin, cold}}<100\, \mathrm{K}$, and the SLED of the warm component should be $\mathrm{L_{CO(13-12)} < L_{CO(12-11)}}$. Without the last constraint, the $T_{\mathrm{kin}}$ of the warm component would be beyond the search range ($>\SI{700}{K}$). 
We add a $1\sigma$ Gaussian noise to the observed CO data and fit the noise-added CO SLED.  This process is repeated 200 times to obtain the mean values and standard deviations. 

\begin{figure}
 \begin{center}
  \includegraphics[width=1.0\linewidth]{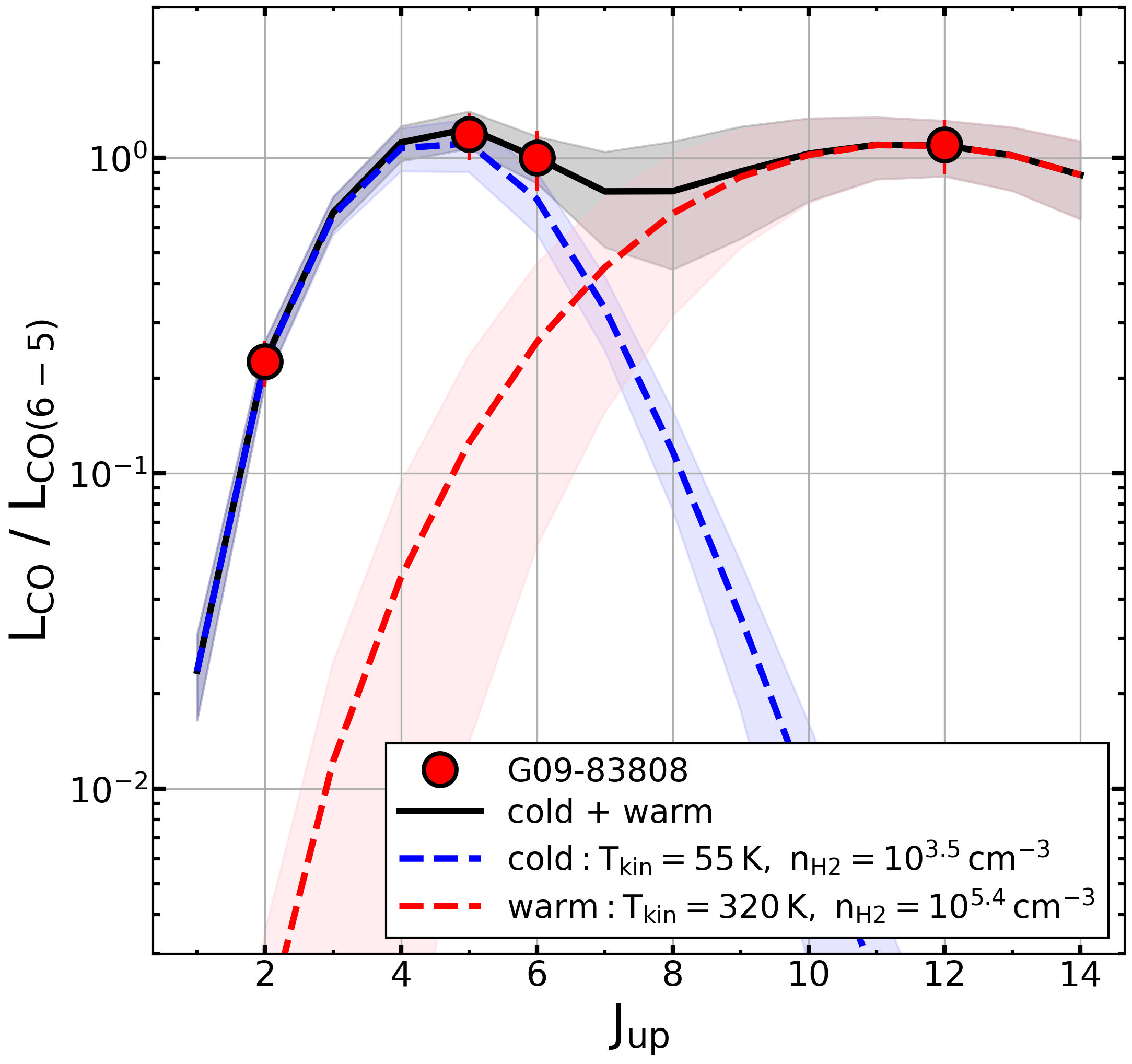}
 \end{center}
\caption{CO SLED from two-component fitting.  The observed data are shown  with red points with error bars.  The contributions from the cold and warm components  are shown with blue and red dashed lines, respectively.  The total  best-fit model is shown with a black line. Color-shaded regions represent the $1\sigma$ uncertainties.}
 \label{CO_RADEX_21_Jy}
\end{figure}

The results are $T_{\textrm{kin}}=55\pm16$\,K, $\log(n_{\mathrm{H2}}/\mathrm{cm^{-3}})=3.5\pm0.5$ for the cold component and $T_{\textrm{kin}}=320\pm170$\,K, $\log(n_{\mathrm{H2}}/\mathrm{cm^{-3}})=5.4\pm0.6$ for the warm component. The temperature of the cold component is consistent with that of the dust estimated from dust spectral energy distribution modeling (51$\:$K; Paper~I). The  results show that more than 99\% of the CO(12-11) emission originates from the warm component (figure~\ref{CO_RADEX_21_Jy}).  
This implies that the warm and dense molecular gas is concentrated in the center of this galaxy ($R_e=0.49^{+0.29}_{-0.19}\ \mathrm{kpc}$).

\begin{table}
\tbl{CO line data (not corrected for lens amplification)}{%
\begin{tabular}{cccc}
\hline\noalign{\vskip3pt} 
Line & $S\Delta V$ (\SI{}{Jy.km.s^{-1}}) & $L$ ($10^8\,L_\odot$) & References\footnotemark[$*$] \\ [2pt] 
\hline\noalign{\vskip3pt} 
CO(2-1)   & $0.6\pm 0.1$   & $0.7\pm 0.1$ & (1) \\
CO(5-4)   & $1.6\pm 0.3$   & $4.8\pm 0.9$ & (2) \\
          & $0.9\pm 0.3$   & $2.8\pm 0.9$ & (3) \\
CO(6-5)   & $0.9\pm0.3$    & $3  \pm 1$   & (2)\\
          & $0.9\pm 0.2$   & $3.1\pm 0.9$ & (3) \\
CO(12-11) & $0.49\pm0.09$  & $3.5\pm 0.7$ & This work\\
\hline\noalign{\vskip3pt} 
\end{tabular}}\label{COdata}
\begin{tabnote}
\hangindent6pt\noindent
\hbox to6pt{\footnotemark[$*$]\hss}\unskip% 
(1) \citealt{Zavala2022}, (2) \citealt{Zavala2018}, (3) \citealt{Fudamoto2017}.
\end{tabnote}
\end{table}

Finally, we discuss the potential effect of dust attenuation, which we  have not considered in  the analysis.
In general, the observed line intensities are affected by dust extinction  in the form: $I=I_{0}\left(1-\mathrm{e}^{-\tau_{\nu}}\right) / \tau_{\nu}$, where $\tau_\nu=(\nu/\nu_0)^\beta$ \citep[e.g.,][]{Rangwala2011}.  Here, the dust optical depth is unity at $\nu=\nu_0(=c/\lambda_0)$ and $\beta$ is the emissivity index.  Adopting $\lambda_0=150\;\micron$ and $\beta=2.5$ (Paper~I),
% performed a dust SED modeling by changing the value of $\lambda_0$ to 100$\:\micron$, 150$\:\micron$, and 200$\:\micron$, and $\lambda_0=\SI{150}{\micron}$ was adopted as the best fit.  
 we find that the effect of dust attenuation on the CO(12-11) line  at the rest-frame wavelength of $\sim\SI{200}{\micron}$ is  roughly 20\%.  If the dust extinction is stronger than expected, the intrinsic brightness of CO(12-11) will be greater, i.e., the temperature of the warm gas will be even higher than the estimated above.
% Since the observed wavelength of CO(12-11) is $\sim\SI{200}{\micron}$, the effect of dust extinction. If the dust extinction is stronger than expected, the intrinsic brightness of CO(12-11) will be greater, i.e., the temperature of the warm gas will be even higher.

% Therefore, when we reconstruct [O\,\emissiontype{III}]\,$88\:\micron$ and [N\,\emissiontype{II}]\,$205\:\micron$ channel maps, we set the number of source objects to be one, and fixed ellipticities (and also position angles) of each sources as 0, corresponding to circle. Thus, free fitting parameters are total flux, position, and effective radius of each sources.  
\section{Discussion and conclusion}

\subsection{Kinematics of ionized gas}\label{sec:kinematic}
Observationally, it remains unclear what causes  extreme starburst in the central compact regions of SMGs, i.e., how the gas is transported to the central compact regions. One possible mechanism  is a gas-rich major merger, which is known from observations and simulations to lead the transportation \citep[e.g.,][]{Mihos1996,Tacconi2008}. 
Another  is a so-called cold stream or minor merger process.  In  this scenario, starburst is caused by the migration into the center of giant star-forming clumps produced by gravitational instability and fragmentation in semi-continuously fed, gas-rich disks \citep[e.g.,][]{Noguchi1999,Dekel2009b,Dekel2009a,Genzel2011,Tadaki2020}.  
In any case, the induced central starburst would build a central bulge component characteristic of compact quiescent galaxies, which are likely to be the post-evolutionary form of SMGs. 

\begin{figure}
 \begin{center}
  \includegraphics[width=1.0\linewidth]{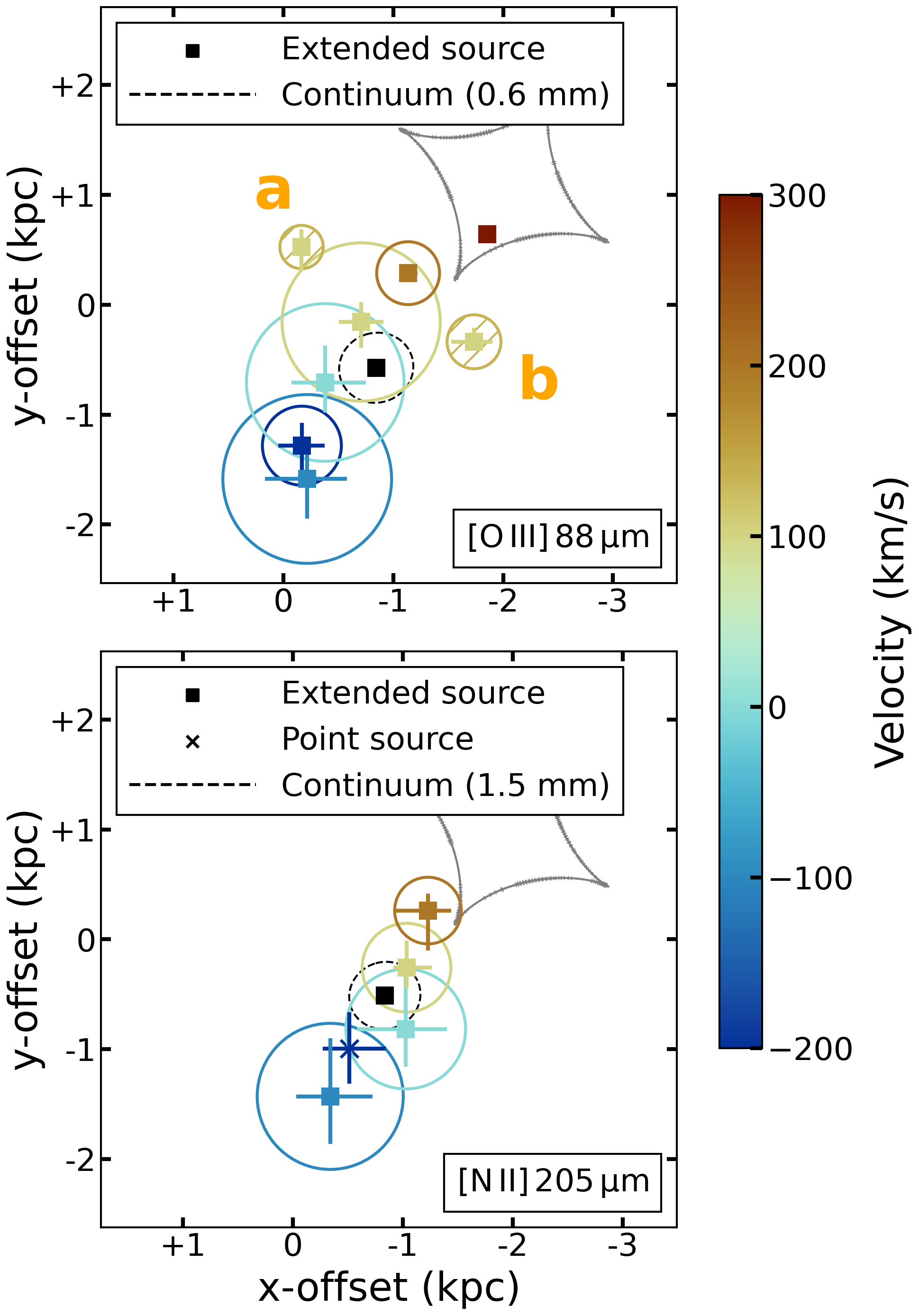}
 \end{center}
\caption{Plot of the reconstructed source positions on the (upper panel) [O\,\emissiontype{III}]\,$88\:\micron$ and (lower) [N\,\emissiontype{II}]\,$205\:\micron$ channel maps.   Filled squares and crosses  indicate the positions of the extended and point sources, respectively.  Circles represent the effective radii.  The reconstructed 0.6$\:$mm and 1.5$\:$mm continuum sources are marked with black squares and dashed ellipses.  Two hatched components (\textbf{a} and \textbf{b}) at [O\,\emissiontype{III}]\,$88\:\micron$ $\SI{0}{km.s^{-1}}$, located along the minor axis of the rotation disk, are the reconstructed sources with two S$\acute{\mathrm{e}}$rsic source models. The caustics of our lens model are shown in gray solid lines.}
 \label{channel_position_1comp}
\end{figure}

% internal process such as gas clump migration into the center due to gravitational instability inside the rotating disk \citep[e.g.,][]{Dekel2009b,Genzel2011}. 
% Another possibility is a internal process such as gas clump migration into the center due to gravitational instability inside the rotating disk \citep[e.g.,][]{Dekel2009b,Genzel2011}. 
The positions and effective radii of the reconstructed sources in each [O\,\emissiontype{III}]\,$88\:\micron$ and [N\,\emissiontype{II}]\,$205\:\micron$ channel are summarized in figure~\ref{channel_position_1comp}. 
% A reconstructed 1.5$\:$mm continuum source is also shown as a dashed ellipse.  
An almost monotonic velocity gradient is found over $\sim 2$\,kpc centered  at the compact dust continuum regions ($R_e\sim0.64$\,kpc; Paper~I)  in both [O\,\emissiontype{III}]\,$88\:\micron$ and [N\,\emissiontype{II}]\,$205\:\micron$ line maps.
And, the reconstructed line profiles of [O\,\emissiontype{III}]\,$88\:\micron$ and [N\,\emissiontype{II}]\,$205\:\micron$ have double-horn shapes as shown in figure \ref{observed_spectrum}.
% Together with the fact that the reconstructed sources from the [N\,\emissiontype{II}]\,$205\:\micron$ integrated intensity image maps are well modeled  with S$\acute{\mathrm{e}}$rsic profiles with  an effective radius of $\sim 1.2$\,kpc and S$\acute{\mathrm{e}}$rsic index $n=1$ (Paper~I), this gradient suggests the existence of an extended rotating ionized gas disk. 
These results could be interpreted as a signature of a rotating disk, but it is difficult to reject the other possibilities such as mergers and outflows with the current data alone due to the limited angular resolution and SNR \citep[e.g.,][]{Simons2019,Rizzo2022}.

Figure~\ref{channel_position_1comp} also shows the two reconstructed sources, designated as sources \textbf{a} and \textbf{b}, where we employ two S$\acute{\mathrm{e}}$rsic sources, at $\SI{100}{km.s^{-1}}$ in the [O\,\emissiontype{III}]\,$88\:\micron$ source map.  Sources \textbf{a} and \textbf{b} are located along the direction perpendicular to the major axis of disk rotation.  These components are not detected in the 0.6$\:$mm, 1.5$\:$mm continuum map and [N\,\emissiontype{II}]\,$205\:\micron$ map. Paper I reported that the 0.6$\:$mm, 1.5$\:$mm continuum map and integrated intensity map of [N\,\emissiontype{II}]\,$205\:\micron$ in G09-83808 is well modeled by a single exponential disk source model, whereas for the integrated intensity map of [O\,\emissiontype{III}]\,$88\:\micron$, some residuals remain in the edges of the two arcs just like in the modeling of [O\,\emissiontype{III}]\,$88\:\micron$ $\SI{100}{km.s^{-1}}$ channel map with a single Gaussian source ($\SI{100}{km.s^{-1}}$ (1 source) row panels in figure \ref{OIII_channel_modeling_1comp}).
Given these facts, these two sources seem to deviate from the primary disk, so we here refer to them as 'sub-components'. 
%  and correspond to (\textbf{a1, a2}) and (\textbf{b1, b2}) components, respectively, in the image plane (figure \ref{OIII_channel_modeling_1comp}).}

%  The SMG G09-83808 has a rotating disk and ionized gas clumps (\textbf{a} and \textbf{b}; see Section~\ref{channelmap}) along the direction perpendicular to the major axis of disk rotation.
%  The clumps are  detected  only with the [O\,\emissiontype{III}]\,$88\:\micron$ line in our observations.
The \textbf{a} and \textbf{b} sub-components account for $13\pm 2\:$\% and $12\pm 2\:$\% of the total [O\,\emissiontype{III}]\,$88\:\micron$ luminosity, respectively.
 From the non-detection ($<3\sigma$) of the sub-components in the dust continuum, the lower limits of the $L_{\mathrm{[OIII]}}/L_{\mathrm{IR}}$ ratios are derived to be $6.4\times 10^{-3}$  and $4.9\times 10^{-3}$ for sub-components \textbf{a} and \textbf{b}, respectively, under the assumption of the same SED as in Paper~I. Oxygen is doubly excited only by hard radiation from  the hottest stars (or AGN). Paper~I showed that the $L_{\mathrm{[OIII]}}/L_{\mathrm{IR}}$ ratio is very sensitive to a variation in the age of star formation, which  alters the energy distribution of incident radiation from the photoionization model. The derived high ratio of $L_{\mathrm{[OIII]}}/L_{\mathrm{IR}}$ indicates that  star formation is going on in these sub-components (see figure~3 of Paper~I). In fact, the high ratio  is consistent with those of local dwarf galaxies ($10^{-3}-10^{-2}$; \citealt{Cormier2015}). 
In massive star-forming galaxies at z=2--4, kpc-scale clumps are gravitationally bounded and are considered to form bulges through some internal process in extended rotating disks \citep[e.g.,][]{Genzel2011,Tadaki2017}.
% which are considered to form bulges  through some internal process in extended rotating disks. 
Also, the presence of star-forming clumps can be explained in the major merger scenario \citep[e.g.,][]{Calabro2019}. 
It is difficult to determine from the current data alone whether the [O\,\emissiontype{III}]\,$88\:\micron$ sub-components are gravitationally bound clumps or have an external origin, such as clumpy gas accretion.

% The presence of  a rotating disk and star-forming clumps are often found in  massive ($M>10^{10}M_\odot$) bulge-forming galaxies at $z\sim 2-4$ \citep[e.g.,][]{Genzel2006,Genzel2008,Tadaki2017,Tadaki2018}, which are considered to form bulges  through some internal process in extended rotating disks.
% based on the detailed analysis on kinematics.  
 % Alternatively, the  presence of a rotating disk and clumps are  explained also in the major merger scenario \citep[e.g.,][]{Robertson2006, Calabro2019}. 
% \textcolor{red}{Just as it is not clear what physical process (rotation or merger or outflow) causes the velocity gradient of [O\,\emissiontype{III}]\,$88\:\micron$ and [N\,\emissiontype{II}]\,$205\:\micron$, the details of the [O\,\emissiontype{III}]\,$88\:\micron$ sub-components, such as whether they are gravitationally bound clumps or whether they are internal to the primary disk are not known from the current data alone.}
 % Hence, it is difficult to  determine from the current data alone which scenario is  the case in G09-83808.  
 % As discussed above, the current data alone cannot reveal the nature of the velocity gradient and [O III] sub-components.
Future observations of bright emissions such as [C \emissiontype{II}]\,$\SI{158}{\micron}$ line with a higher angular resolution will enable us to quantitatively investigate the kinematics of the multiple phases of the ISM and nature of the sub-components.

\subsection{Heating source of CO(12-11) emission}
\begin{figure}
 \begin{center}
  \includegraphics[width=1.0\linewidth]{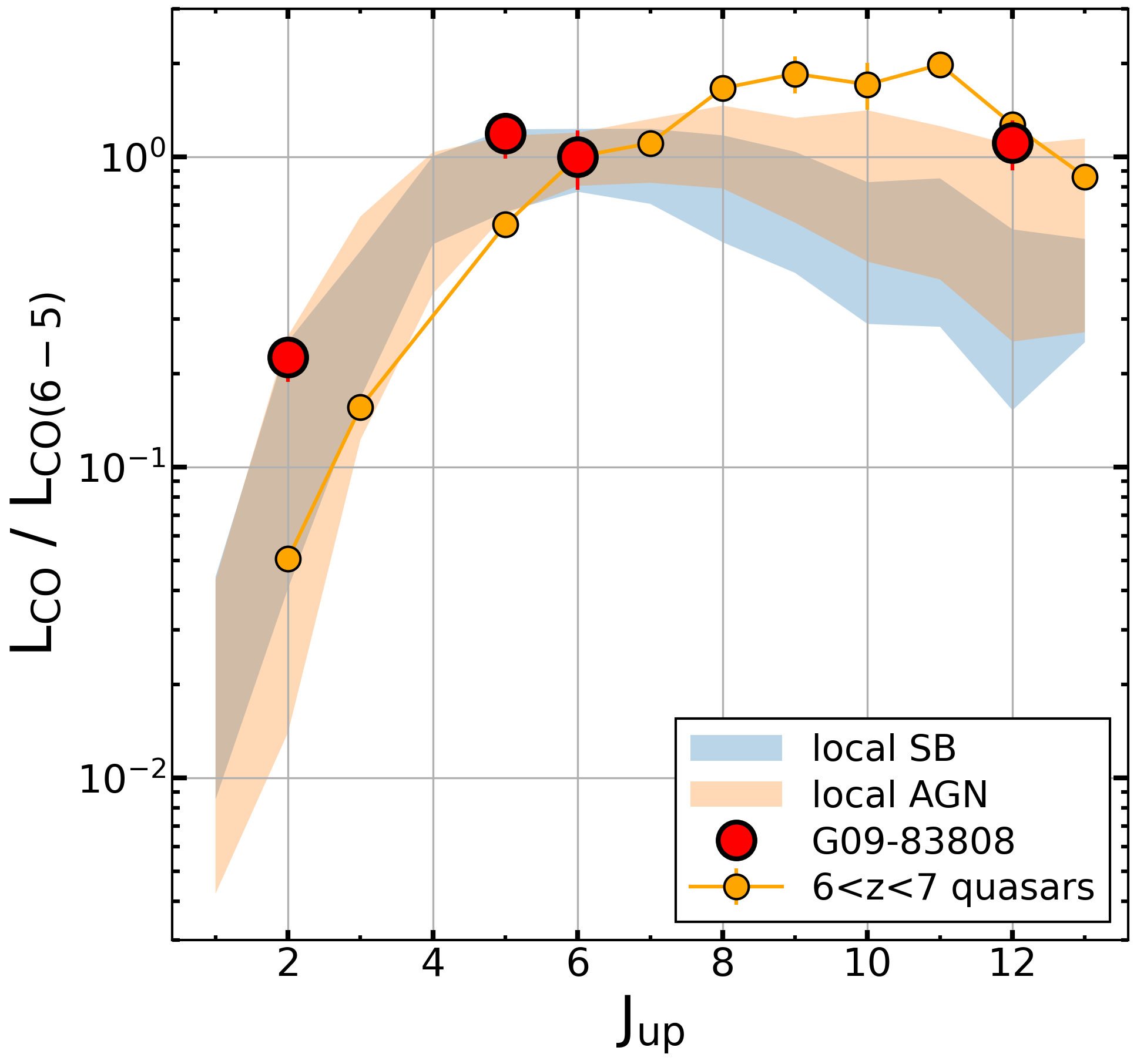}
 \end{center}
\caption{CO(6-5) normalized CO SLED of G09-83808 and other systems (local starburst galaxies, local AGNs, and $6<z<7$ quasars) for comparison.  The data points of the local starburst galaxies and local AGNs are shown  with shaded regions in sky-blue and orange, respectively \citep{Rosenberg2015, Mashian2015}.   That of the $6<z<7$ quasars is also shown in orange \citep{Carniani2019, Li2020}.} % Original(WB): \citealt{Bothwell2013}. \citealt{Carniani2019} and \citealt{Li2020}.
\label{CO_SLED}
\end{figure}

Detection of high-$J$ CO lines suggests the existence of high-energy heating sources.   
The CO lines from $J_\mathrm{{up}} = 5$ to $J_\mathrm{{up}} = 13$ in the nearest AGN, Mrk~231, were studied  with Herschel observations \citep{vanderwerf2010}.  
They argued that the measured CO($J > 10$) luminosities could not be explained  with dense photodissociation-region models and that X-rays from the accreting supermassive black hole (SMBH) dominantly caused the excitation of the CO($J > 10$) emission.  Furthermore, CO($J > 10$) emission has been  detected in quasars at $z = 6-7$ \citep{Gallerani2014,Wang2019,Li2020} and is considered to be radiated from X-ray-dominated regions around AGNs.  
Figure~\ref{CO_SLED} shows CO(6-5)-normalized CO SLEDs of G09-83808, local starburst galaxies, local AGNs \citep{Mashian2015, Rosenberg2015}, and average of 4 quasars  at $6<z<7$ \citep{Carniani2019,Li2020}.  The high ratio of  $L_\mathrm{{CO(12-11)}} / L_\mathrm{{CO(6-5)}}=1.1\pm0.2$ in G09-83808  agrees better with those in local AGNs and $6<z<7$ quasars than those in local starburst galaxies.  This  comparison suggests that G09-83808 hosts a dust-obscured AGN in the center although we cannot  exclude other possibilities, such as shock \citep[e.g.,][]{Meijerink2013, Saito2017}  and extreme starburst \citep[e.g.,][]{Riechers2013}, with the current data alone.
% Although there are other excitation mechanisms such as shock \citep[e.g.,][]{Meijerink2013} or extreme starburst \citep[e.g.,][]{Riechers2013}, G09-83808 could be one of the first dust obscured AGNs in the epoch of reionization.  
One powerful tool to explore the dust-obscured AGNs is polycyclic aromatic hydrocarbon emission features in the rest-frame mid-infrared spectrum because small PAHs would be destroyed by hard X-rays from AGN \citep[e.g.,][]{Moorwood1986, Imanishi2000}. Mid Infrared Instrument on \textit{James Webb Space Telescope} will enable us to observe these lines.  Exploration of dust-obscured AGNs will  help us reveal the nature of the early co-evolution of SMBHs and host galaxies.

% Theoretical simulations suggest the existence of dust obscured AGNs as precursors of QSOs in SMGs \citep[e.g.,][]{Hickox2018,Ni2020}, and also observational signs has been found \citep[e.g.,][]{Davies2019}. G09-83808 which has high CO excitation similar to that of the local and high-z QSO-host galaxies is a good target for studying the presence of dust obscured AGNs. Thus, future ALMA observations of higher-J CO lines (i.e., $J_{\mathrm{up}} \geq 14$) are needed to distinguish whether the bright high-$J$ CO emissions are excited by intense star formation or powerful AGN.  The polycyclic aromatic hydrocarbon (PAH) emission features is also a powerful tool for separating these kinds of activity and estimating their relative importance because the features are associated only with starburst activity and not with AGN activity \citep{Moorwood1986}.  The PAH emission line at $\lambda_{\textrm{rest}}=\SI{3.3}{\micron}$ is redshifted to $\SI{23}{\micron}$ which is observable with Mid Infrared Instrument (MIRI) on James Webb Space Telescope. 

\begin{ack}
We thank the referee for a number of constructive comments and suggestions that improved the paper. We wish to thank Masamune Oguri and Justin S. Spilker for lots of advice about gravitational lens modeling.  This paper makes use of the following ALMA data: ADS/JAO.ALMA\#2019.1.01307.S. ALMA is a partnership of ESO (representing its member states), NSF (USA) and NINS (Japan), together with NRC (Canada), MOST and ASIAA (Taiwan), and KASI (Republic of Korea), in cooperation with the Republic of Chile. The Joint ALMA Observatory is operated by ESO, AUI/NRAO and NAOJ. We thank the ALMA staff and in particular the EA-ARC staff for their support. This work was supported by JSPS KAKENHI Grant Numbers 20K14526 and 17H06130, and by the NAOJ ALMA Scientific Research Grant Number 2017-06B. Data analysis was in part carried out on the Multiwavelength Data Analysis System operated by the Astronomy Data Center (ADC), National Astronomical Observatory of Japan (NAOJ).
\end{ack}

\begin{appendix}
\section{CO(12-11) channel maps}
\begin{figure*}
 \begin{center}
  \includegraphics[width=1.0\linewidth]{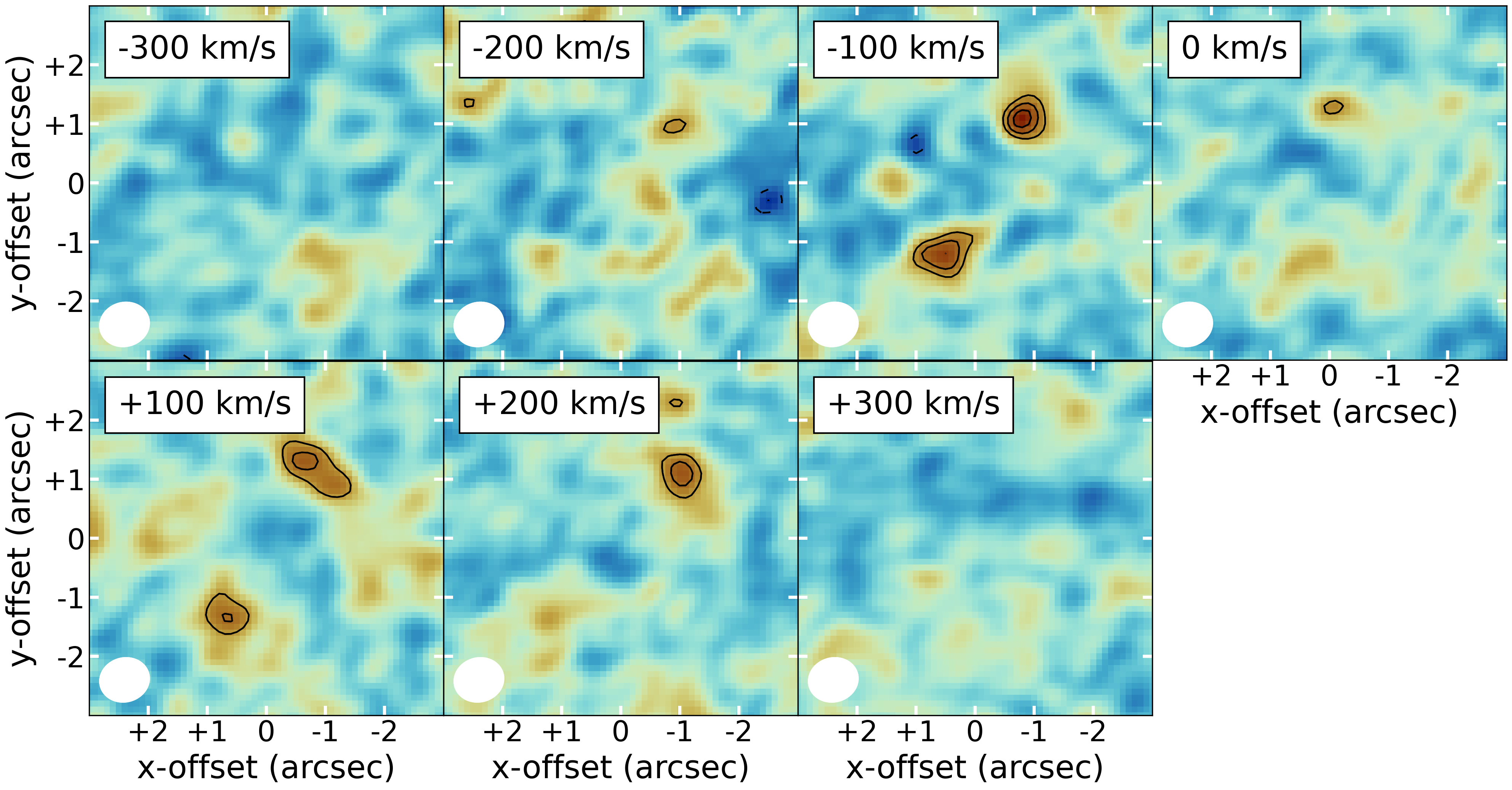}
 \end{center}
\caption{Cleaned CO(12-11) channel maps from $\SI{-300}{km.s^{-1}}$ to $\SI{300}{km.s^{-1}}$ with robust parameter of 2.0. The peak SNRs are [$2.7\sigma$, $3.4\sigma$, $5.7\sigma$, $3.4\sigma$, $4.4\sigma$, $4.6\sigma$, $2.5\sigma$]. The black contours are drawn at [3, 4, 5]$\times \sigma$ where $1\sigma=\SI{0.11}{mJy.beam^{-1}}$. The synthesized beam size is displayed in the lower-left corner.}
\label{CO1211_channelmaps}
\end{figure*}

Figure \ref{CO1211_channelmaps} shows the cleaned CO(12-11) channel maps with a robust parameter of 2.0. Due to the low SNR, we didn't perform the source reconstructions for each channel.

\section{CO(12-11) intrinsic source size}
In the section \ref{co1211mom0}, the CO(12-11) intrinsic source size is derived to be $R_e=0.08^{+0.05}_{-0.03}\ \mathrm{arcsec}$ with the help of lensing magnification ($\mu_{\mathrm{CO(12-11)}}=9.3^{+7.2}_{-1.7}$). However, since the angular resolution is 0\farcs87$\times$0\farcs77, the effective angular resolution in the source plane is $0\farcs8/\sqrt{\mu}\sim 0\farcs26$, corresponding to the disks with a radius of $\sim0\farcs13$. This value is $\sim1.6$ times larger than the derived source size, so it seems strange that the source size is measured despite the lack of angular resolution.  
There are two possible explanations for the reason why \texttt{GLAFIC} can estimate a source size smaller than the beam size. 

First, since the beam profile is known, in principle, the size of objects smaller than the beam size can be measured (if not too small). 

\begin{figure}
 \begin{center}
  \includegraphics[width=1.0\linewidth]{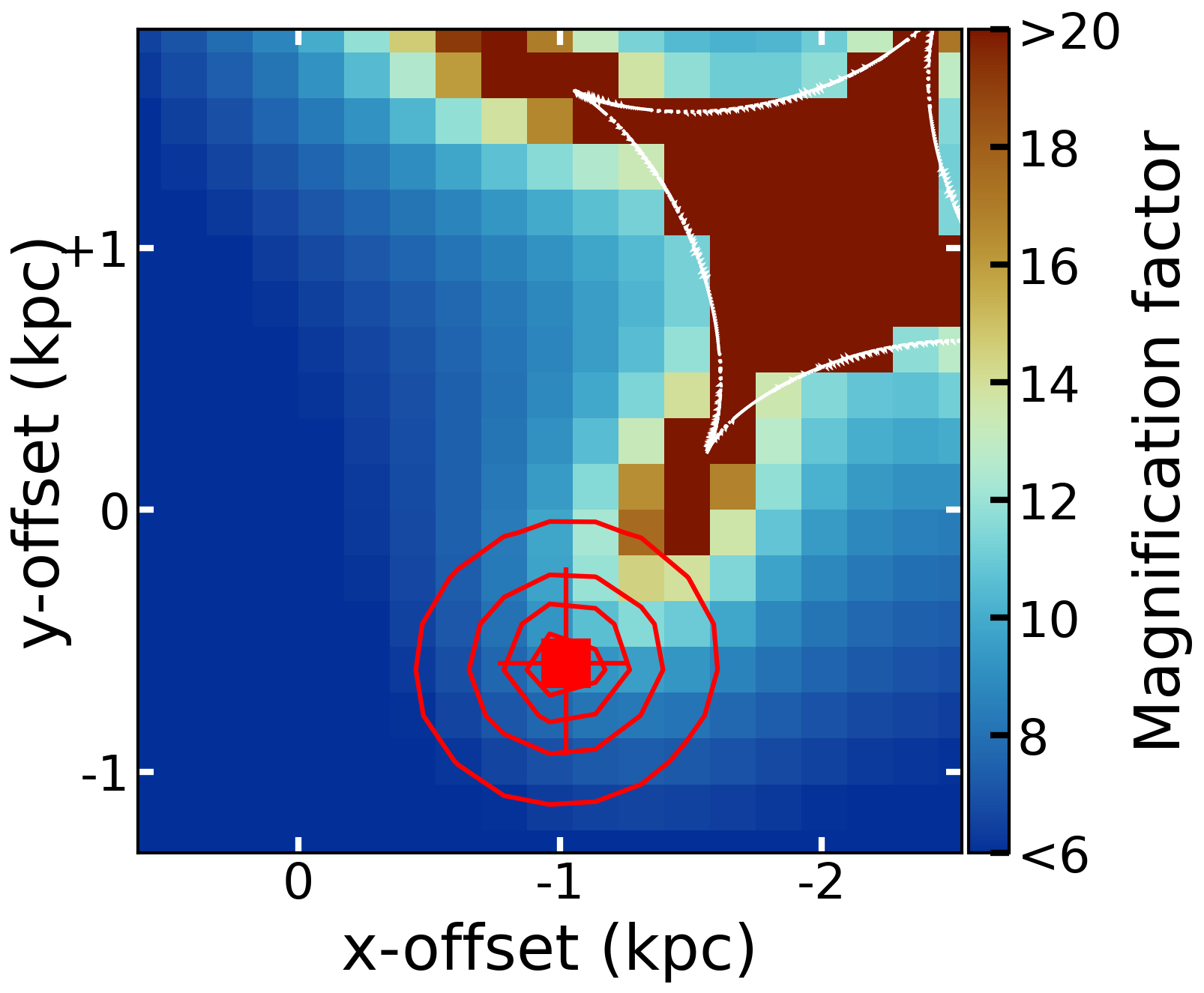}
 \end{center}
\caption{Distribution of the magnification factor in the source plane at $z=6.0255$. The red contours represent the 20\%, 40\%, 60\%, 80\% of the peak of the reconstructed CO(12-11) integrated intensity map. The filled red square and cross indicate the peak position and its uncertainty (see section \ref{co1211mom0}). The caustics of our lens model is shown as the white solid line.}
\label{CO1211_magmap}
\end{figure}

Second, since we assume a source profile (S$\acute{\mathrm{e}}$rsic profile with an index of $n=1$) in this modeling, we do not necessarily need to directly resolve the object to measure its size. 
For example, if the profile is fixed, it would be possible to regulate the source size by extrapolating the source profile to the center using information from the fitting of the outskirt parts and the central flux, and this may be what is happening. 
In fact, the magnification map shown in figure \ref{CO1211_magmap}, shows that the outskirt parts near the caustics are magnified as high as about 15-20 times, although the average magnification factor is $\sim9.3$.
Regarding the second point, fitting with the source profile free would provide more robust results. Therefore, we perform the fitting with the S$\acute{\mathrm{e}}$rsic index also free. Consequently, the obtained effective radius don’t change significantly ($R_e=0.09^{+0.07}_{-0.04}\ \mathrm{arcsec}$, $n=0.6^{+3.3}_{-0.06}$). 

In addition, we try the source reconstructions using clean images with different robust parameters of $1.5$, $1.0$, $0.5$, $0.0$, and $-0.5$ (the same as that of dust emission). 
When imaging with robust parameters of $0.0$ or less, the integrated flux density is less than $5\sigma$, so we don't perform the modeling. With the parameter of 0.5, the resolution is 0\farcs66$\times$0\farcs59 and the source size is $R_e=0.08^{+0.05}_{-0.03}$ arcsec, which is consistent with that of the parameter of $2.0$. 

\begin{figure}
 \begin{center}
  \includegraphics[width=1.0\linewidth]{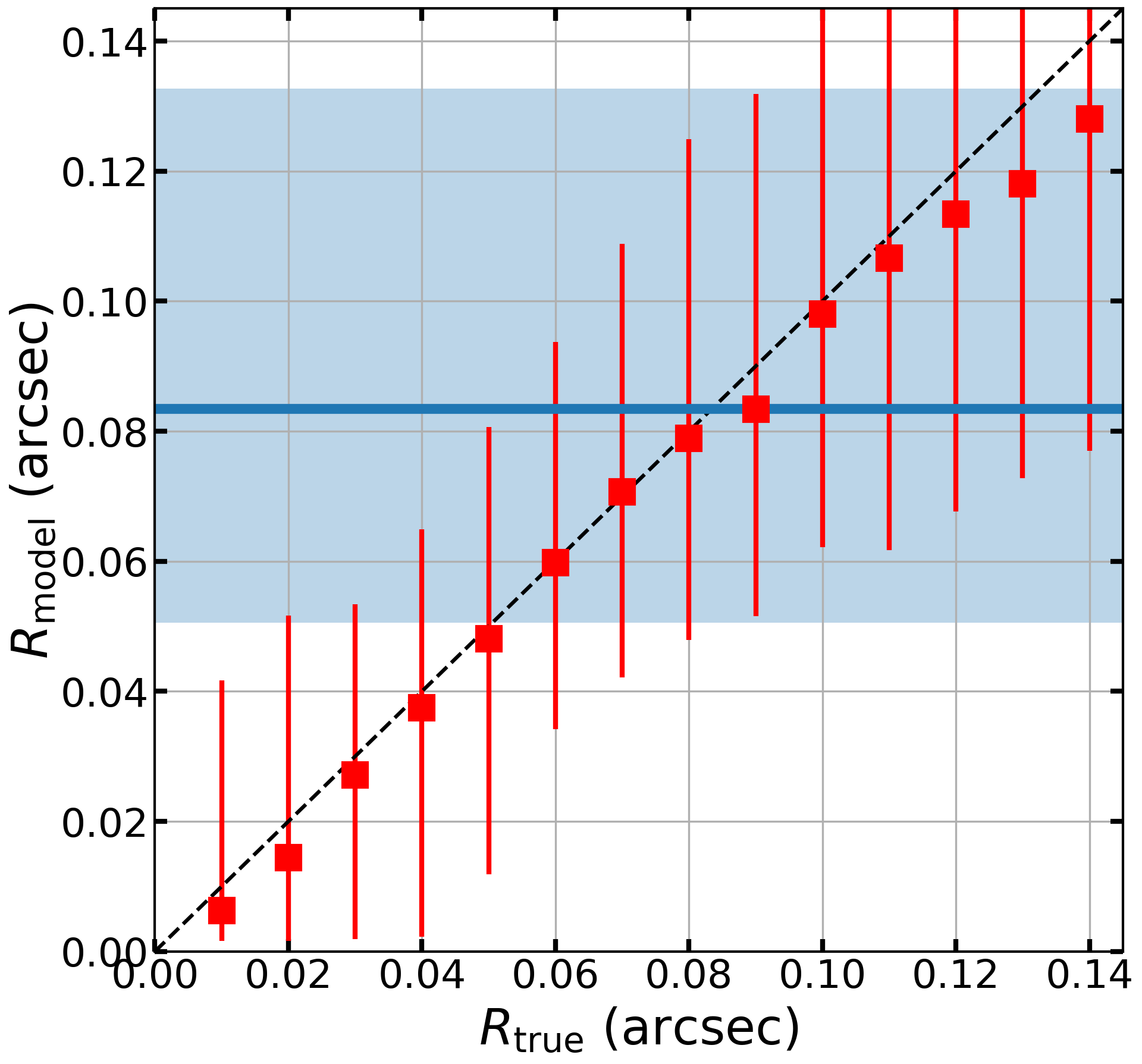}
 \end{center}
\caption{Model vs. true effective radius for intrinsic CO(12-11) emission. The blue line and shaded region represent the estimated CO(12-11) intrinsic source size ($R_e=0.08^{+0.05}_{-0.03}\:\mathrm{arcsec}$).
The red squared dots are the median values of measured effective radii with $1\sigma$ uncertainties from 500 Monte Carlo runs.
The black dashed line denotes the relation $R_{\text{true}}=R_{\text{model}}$.}
\label{CO1211_simulation}
\end{figure}

Finally, we simulate the source size measurements by making mock images. 
Using \texttt{GLAFIC} code, we vary the effective radius of the S$\acute{\mathrm{e}}$rsic source from 0\farcs01 to 0\farcs14 in 0\farcs01 increments while fixing the other parameters at the best-fit parameters obtained in section \ref{co1211mom0} to create the mock lensed images.
We add the $1\sigma$ noise map convolved with a dirty beam to the images and perform the source reconstructions for the noise-added mock images in the same manner as section \ref{co1211mom0}. Then, we take the median values, 16th and the 84th percentile of the best-fit values from 500 Monte Carlo runs as the best-fit measured sizes and uncertainties, respectively.
In figure \ref{CO1211_simulation}, the measured effective radius ($R_{\text{model}}$) is plotted as a function of the true effective radius ($R_{\text{true}}$). This result implies that it is difficult to distinguish the extended source from the point source in the region of $R_{\text{true}}\lesssim 0.04\:$arcsec. However, when the $R_{\text{model}}\sim 0.08\:$arcsec, it is possible to measure the source size as the extended source.
% Finally, when the fit was performed using a point source, the reduced chi-square value was 1.85 and 1.53 for the parameters of 2.0 and 0.5, respectively, which was worse than when an extended S$\acute{\mathrm{e}}$rsic source was used ($1.26$ and $1.17$).  

From these results, we have concluded that the intrinsic source size of CO(12-11) emission is measured as an extended source, not a point source. 
Note that we used the measured size when the S$\acute{\mathrm{e}}$rsic index is fixed to 1.0, in order to make the size comparison with other emissions fairer.

\end{appendix}

% \section{Case of two or more paragraphs}
% \section{Case of two or more paragraphs}

%%%
% See the manual for the detail.
%%%

\bibliographystyle{apj}
\bibliography{MyCollection.bib}
\end{document}